\documentclass{ifacconf}

\usepackage{graphicx}     
\usepackage{natbib}        
\usepackage{amsmath,amssymb}
\usepackage{mathrsfs}
\usepackage{color}
\theoremstyle{definition}

\newcommand{\Ltwo}{\boldsymbol{\rm L}_{2}}
 
\newcommand{\Ltwoe}{\boldsymbol{\rm L}_{2e}} 
\begin{document}
\begin{frontmatter}

\title{On output consensus of heterogeneous dynamical networks \thanksref{footnoteinfo}} 

\thanks[footnoteinfo]{This work was supported in part by the National Science and Technology Council of Taiwan under No. 113-2222E-110-002-MY3}

\author[First]{Yongkang Su} 
\author[First]{Lanlan Su} 
\author[Third]{Sei Zhen Khong}

\address[First]{Department of Automatic Control \& Systems Engineering, University of Sheffield, Sheffield, UK (e-mail: ysu34@sheffield.ac.uk \& lanlan.su@sheffield.ac.uk)}
\address[Third]{Department of Electrical Engineering, National Sun Yat-sen University, Kaohsiung 80424, Taiwan (e-mail: szkhong@mail.nsysu.edu.tw)}

\begin{abstract}                
This work is concerned with interconnected networks with non-identical subsystems. We investigate the output consensus of the network where the dynamics are subject to external disturbance and/or reference input. For a network of output-feedback passive subsystems, we first introduce an index that characterises the gap  between a pair of  adjacent subsystems by the difference of their input-output trajectories. The set of these indices  quantifies the level of heterogeneity of the networks. We then provide a condition in terms of the level of heterogeneity and the connectivity of the networks  for ensuring the output consensus of  the interconnected network.
\end{abstract}


\end{frontmatter}

\section{Introduction}
Over the past few decades, consensus control of interconnected networks has received a wide range of research interests and extensive applications in many areas, such as robot coordination \cite[]{qiao2015consensus}, power grid \cite[]{yang2013consensus}, and distributed sensor networks \cite[]{olfati2005consensus}. In particular, passivity-based approaches have outstanding relevance in the consensus analysis for interconnected networks, and fruitful research results have been achieved, see, e.g., \cite{chopra2006passivity,burger2015dynamic}. For example, \cite{chopra2006passivity} studied the output consensus of passive multi-agent systems over weight-balanced digraphs. In addition, by the internal model approach, the consensus problem for a network of incrementally passive systems over dynamic diffusive coupling was investigated in \cite{burger2015dynamic}.

It should be noted that a common feature of the aforementioned literature is that the subsystems are expected to be passive. In the engineering practice, however, many systems are not inherently passive, \cite[]{kelkar1998robust}. Recently, there are fruitful results focusing on the consensus problem of interconnected networks with non-passive subsystems, see, e.g., \cite{stan2007output,qu2014modularized,zhang2018cooperative,li2019consensus}. For example, \cite{stan2007output} studied the consensus problem for networks of cyclic biochemical oscillators with identical incrementally output-feedback passive systems. Besides, the consensus problem of multi-agent systems with input feedforward passive agents over diffusive coupling was investigated in \cite{li2019consensus}.

Moreover, in \cite{scardovi2010synchronization}, the consensus problem of interconnected networks with incrementally output-feedback passive systems and external inputs was studied from a purely input-output perspective. A condition of output with a high level of consensus is provided by combining the input-output properties of the subsystems with the connectivity of the network. However, it focused only on homogeneous networks, i.e., the dynamics of the subsystems in the interconnected networks are identical, which might be restrictive and impractical in many cases. In engineering practice, all physical systems of the interconnected systems are not exactly identical due to certain undesirable environmental factors and parametric uncertainties \cite[]{li2014distributed}. Therefore, this work attempts to generalise the research result to the case of heterogeneous networks.

The main contributions of this work are summarized as follows: 1) An index that characterises the gap between a pair of adjacent subsystems is introduced by the difference of their input-output trajectories, and the set of these indices quantifies the level of heterogeneity of the networks. 2) A condition in relation to the output consensus of the heterogeneous network is proposed in terms of the level of heterogeneity and the connectivity of the networks.

\section{Preliminaries}
\subsection{Notation}
Let $\mathbb{R}$ be the set of real numbers. For a matrix $A$, denote by $A^T$ its transpose, and $\text{rank}(A)$ its \text{rank}. Denote by $I_m$ the $m \times m$ identity matrix. Let $\textbf{1}_m := [1, \ldots ,1]^T \in {\mathbb{R}^m}$. Given scalars ${{a_1}, \ldots ,{a_m}}$, let the column vector ${\rm col}\left( {{a_1}, \ldots ,{a_m}} \right) := {\left[ {a_1, \ldots ,a_m} \right]^T}$ and $\text{diag}\{a_1,\dots, a_m\}$ the diagonal matrix with its $i$th diagonal entry being $a_i$. Given a symmetric matrix $A=A^T$, we use $A\succ 0$ (resp., $A\succcurlyeq 0$) to denote that $A$ is positive definite (resp., positive semi-definite). 
Define the signal space $\Ltwo =\left\{ {x:\left[ {0,\infty } \right) \to \mathbb{R}^m|{{\left\| x \right\|}^2}: = \int_0^\infty  {{{\left| {x(t)} \right|}^2}} } \right.{\hspace{-2mm}}\left. {dt < \infty } \right\}$ where $|\cdot|$ denotes the Euclidean norm.
For any $x:\left[ {0,\infty } \right) \to {\mathbb{R}^m}$, define the truncation operator $\left( {{P_T}x} \right)(t) = x(t)$ for $t \le T$ and $\left( {{P_T}x} \right)(t) = 0$ for $t>T$. Define $\Ltwoe$ as $\Ltwoe=\left\{ {x:\left[ {0,\infty } \right) \to {\mathbb{R}^m}{\hspace{0.3mm}}|{\hspace{0.3mm}}{P_T}x \in \Ltwo ,\forall T \ge 0} \right\}$. 
Given $x \in \Ltwoe$ and $ T \ge 0$, ${\left\| x \right\|_T} := {\left( {\int_0^T {{{\left| {x(t)} \right|}^2}dt} } \right)^{\frac{1}{2}}}$. Given $x,y \in \Ltwoe $ and $T \ge 0$, ${\left\langle {x,y} \right\rangle _T} := \int_0^T {{x^T}(t)y(t)dt}$. An operator $H:\Ltwoe \to \Ltwoe$ is said to be causal if ${P_T}H{P_T} = {P_T}H,\,\forall T \in \mathbb{R}$. 

\subsection{Graph Theory}\label{subsetion_Graph_Theory}
A graph is defined by $\mathcal{G} = (\mathcal{N},\mathcal{E} )$, where $\mathcal{N} = \{ 1, \ldots ,n\} $ is the set of nodes and $\mathcal{E} \subset \mathcal{N} \times \mathcal{N}$ is the set of edges. The edge $(i,j) \in \mathcal{E}$ denotes that node $i$ can obtain information from node $j$. Let $\mathcal{N}_i=\left\{ {j \in \mathcal{N}{\hspace{0.8mm}}|{\hspace{0.8mm}}(i,j) \in \mathcal{E}} \right\}$ denote the set of neighbours of node $i$. The graph $\mathcal{G}$ is said to be undirected if $(i,j) \in \mathcal{E}$ then $(j,i) \in \mathcal{E}$. $\mathcal{G}$ is said to be strongly connected if there exists a sequence of edges between every pair of nodes. For a graph $\mathcal{G}$, its adjacency matrix $\mathcal{A}=[a_{ij}]\in\mathbb{R}^{n\times n}$ is defined as $a_{ij}=1$ if $(i,j)\in \mathcal{E}$ and $a_{ij}=0$ otherwise. 
It is assumed that there are no self-loop, that is $a_{ii}=0, i=1, \dots, n$. The Laplacian matrix $L=[l_{ij}]\in\mathbb{R}^{n\times n}$ of $\mathcal{G}$ is defined as: 
\begin{equation*}
{l_{ij}} = \left\{ {\begin{array}{*{20}{c}}
{ \sum\limits_{q = 1}^n {{a_{iq}}}, i=j}\\
{-a_{ij},i \ne j}.
\end{array}} \right.
\end{equation*} 
For an undirected graph $\mathcal{G}$, we may assign an orientation to $\mathcal{G}$ by considering one of the two nodes of an edge to be the positive end and the other one to be the negative end. Denote by $\mathscr{L}_i^ +$ (resp., $\mathscr{L}_i^ -$) the set of edges for which node $i$ is the positive (resp. negative) end. Let $p$ be the cardinality $\mathcal{E}$, i.e., the total number of edges. Define the incidence matrix $D=[d_{ik}]\in\mathbb{R}^{n\times p}$ of an undirected graph $\mathcal{G}$ as
\begin{equation*}
{d_{ik}} = \left\{ 
\begin{matrix}
     + 1, & k \in \mathscr{L}_i^ + \\
 - 1, & k \in \mathscr{L}_i^ - \\
0,& \mathrm{otherwise}.
\end{matrix}
\right.
\end{equation*}
For an undirected graph $\mathcal{G}$, it holds that $D^T \textbf{1}_n =0 $ and $L=DD^T$ \cite[Definition 1.2]{bai2011cooperative}. A spanning tree in $\mathcal{G}$ is an edge-subgraph of $\mathcal{G}$ which has $n-1$ edges and contains all nodes \cite[p29]{biggs1993algebraic}.

\subsection{Passivity}
In this work, we adopt the definitions of passivity from Definition 2.2.1 in \cite{van2000l2} and incremental output-feedback passivity from Definition 2 in \cite{scardovi2010synchronization} for system described by input-output maps.
\begin{defn}\label{def: passive}
A causal operator $H:\Ltwoe \to \Ltwoe $ is said to be passive if there exists some constant $\beta \in \mathbb{R}$ such that for all $u \in \Ltwoe$
\begin{equation}\label{eq: passive}
 {\left\langle {Hu,u } \right\rangle _T}\ge \beta,\,\forall T \ge 0,
\end{equation}
and $\gamma$-incrementally output-feedback passive (OFP) if there exist $\gamma \in \mathbb{R}$ and $\beta \in \mathbb{R}$ such that for all $u,v \in \Ltwoe$
\begin{align}\label{eq: incrementally OFP passive}
 {\left\langle {Hu-Hv,u-v } \right\rangle _T}\ge \gamma\left\| {Hu - Hv} \right\|_T^2+\beta,\,\forall T \ge 0.
\end{align}
\end{defn}
\subsection*{Problem Formulation}\label{sec: problem formulation}
Consider a group of $n$ systems $H_i:\Ltwoe \to\Ltwoe$ described by
\begin{equation}\label{eq: system model}
    {y_i} = {H_i}{u_i}, \,i\in\{1,2,\ldots,n\},
\end{equation}
where $u_i, y_i \in \Ltwoe$ denote respectively the input and output of the $i$-th system. Suppose the group of systems is interconnected by means of an undirected and connected graph $\mathcal{G} = (\mathcal{N},\mathcal{E})$. Specifically, the input $u_i$ to the $i$-th system, is given by
\begin{equation}\label{eq: input}
u_i= w_i-v_i,\,  i \in \mathcal{N}.
\end{equation}
Here, $w_i \in \Ltwoe$ is the external disturbance and/or reference signals, and $v_i \in \Ltwoe$ depends on the relative outputs between the $i$-th system and its neighbours as given by
\begin{equation}\label{eq: coupling}
v_i = \sum\limits_{j \in \mathcal{N}_i} {{\alpha _{ij}}\left( {{y_i} - {y_j}} \right)},
\end{equation}
where the scalars $\alpha_{ij}=\alpha_{ji} >0$. Let $Y := {\rm col}\left( {{y_1}, \ldots ,{y_n}} \right)$ and the same notation is used to define the vectors $V$, $W$ and $U$. 
Substituting \eqref{eq: coupling} into \eqref{eq: input} and recalling the definition of the incidence matrix $D$, it can be obtained that
\begin{equation}\label{eq: iuput vector}
U = W - V = W - D\Psi D^T Y,
\end{equation}
where $\Psi = \text{diag}\{\alpha_1,\dots, \alpha_p \}$ with ${\alpha _k} = {\alpha _{ij}}, k \in \{1, \dots , p\}$ if $d_{ik}=1$ and $d_{jk}=-1$.

The aim of this work is to derive conditions such that there exists a gain $\rho>0$ and a constant $\varepsilon\ge 0$ such that  
\begin{align}\label{eq:aim}
    {\left\| {D^T Y} \right\|_T}\le \rho{\left\| {D^T W} \right\|_T}+\varepsilon,\,\forall {W } \in \Ltwoe,\,\forall T \ge 0.
\end{align}

Note that the external input $W$ can be considered to be the sum $W=W_1+W_2$ with $W_1=w\mathbf{1}_n, w\in \Ltwoe$ being a reference signal and $W_2\in\Ltwo$ being disturbance. Then, if \eqref{eq:aim} holds, it implies that
${\left\| {D^T Y} \right\|}\le \rho{\left\| {D^T W_2} \right\|}+\varepsilon$. As remarked in  \cite{scardovi2010synchronization}, ${\left\| {D^T Y} \right\|_T}$ quantifies the synchrony of the outputs in the time interval $[0,T]$, and \eqref{eq:aim} implies that the interconnected network enjoy the property that external input with a high level of consensus produces output with the same property. More importantly, \eqref{eq:aim} can be extended to ensure synchronisation in systems
described with a state space formalism (with arbitrary initial conditions) under the assumption of zero-state reachability. 

\section{Main Result}
Given two systems  $H_i$ and $H_j$, suppose they are $\gamma$-incrementally output-feedback passive, i.e.,
$$ {\left\langle {{H_i}u - {H_i} v,u - v} \right\rangle _T} \ge {\gamma}\left\| {{H_i}u - {H_i} v} \right\|_T^2+\beta_i$$
and
$$ {\left\langle {{H_j}u - {H_j} v,u - v} \right\rangle _T} \ge {\gamma}\left\| {{H_j}u - {H_j} v} \right\|_T^2+\beta_j$$
for all $u,v \in \Ltwoe$ and $T \ge 0$. We introduce in the next assumption an index  $\gamma_{ij}$ to characterise the gap between $H_i$ and $H_j$.  
\begin{assum}\label{assump: 1} 
For all $T \ge 0$, there exist $\gamma_{ij}\in\mathbb{R} $ and $ \beta_{ij}\in\mathbb{R}$ with $(i,j) \in \mathcal{E}$ such that the operators ${H_i}$ and $H_j$ satisfy
\begin{align} \label{eq: assump 1}
 {\left\langle {{H_i}u - {H_j} v,u - v} \right\rangle _T} {\hspace{-0.5mm}}\ge{\hspace{-0.5mm}} {\gamma _{ij}}\left\| {{H_i}u {\hspace{-0.5mm}}-{\hspace{-0.5mm}} {H_j} v} \right\|_T^2+\beta_{ij},\forall u,v \in \Ltwoe.   
\end{align}
\end{assum}\par

\begin{rem}
When $H_i=H_j$, \eqref{eq: assump 1} reduces to $H_i,H_j$ being $\gamma_{ij}$-incrementally output feedback passive. The deviation of $\gamma_{ij}$ from $\gamma$ capture the gap between $H_i$ and $H_j$.
\end{rem}

Given an undirected and connected graph $\mathcal{G}$ with an incidence matrix $D\in\mathbb{R}^{n\times p}$,  let $\mathcal{G}_{ST}$ be any spanning tree of $\mathcal{G}$ and let $D_{ST}\in\mathbb{R}^{n\times (n-1)}$ be the incidence matrix of $\mathcal{G}_{ST}$. We present next two supporting lemmas. 

\begin{lem}\label{lem: Q}
    There exists a matrix $Q \in \mathbb{R}^{(n-1) \times p}$ such that 
 $D=D_{ST}Q$ and $\text{rank}(Q)=n-1$.
\end{lem}
\begin{pf}
Since $\mathcal{G}$ is an undirected and connected graph and $\mathcal{G}_{ST}$ is a spanning tree of $\mathcal{G}$, $\text{rank}(L) = \text{rank}(D^TD)=\text{rank}(D)=\text{rank}(D_{ST}) = n-1$ and there must exist a matrix $Q \in \mathbb{R}^{(n-1) \times p}$ such that $D=D_{ST}Q$. Noting that $\text{rank}(AB) \le \text{min}\{\text{rank}(A),\text{rank}(B)\}$, one has $\text{rank}(Q)=n-1$.\hfill $\qed$
\end{pf}

\begin{lem}\label{lem: positive definite}
Given $Q\in\mathbb{R}^{(n-1)\times p}$  such that $D=D_{ST}Q$,  $R =\text{diag}\{r_1,\ldots,r_p\} $ with $r_i \ge r >0, i\in \{1, \dots, p\}$, and $\gamma \in \mathbb{R}$.
It holds that  
 \begin{align*}
     M := Q\left( {{\gamma}{I_p}R + {R}{D^T}DR} \right){Q^T}\succ 0
 \end{align*}
 if  $\gamma + r \lambda_2 >0$ where $\lambda_2$ is the second smallest eigenvalue of the Laplacian matrix $L$.
\end{lem}
\begin{pf}
Recall from Lemma \ref{lem: Q} that $\text{rank}(Q)=n-1$. By performing singular value decomposition \cite[Theorem 1.11]{dullerud2013course}, we can write ${R^{\frac{1}{2}}}{D^T}D{R^{\frac{1}{2}}} = V\left[ {\begin{array}{*{20}{c}}
\Sigma &{}\\
{}&0
\end{array}} \right]{V^T}$, where $V \in \mathbb{R}^{p \times p}$ is a unitary matrix, $\Sigma  = \text{diag}\left\{ {\theta _1^2, \ldots ,\theta _{n - 1}^2} \right\}$ with ${\theta _1} \ge  \cdots  \ge {\theta _{n - 1}} > 0$, and ${\theta _1} , \dots , {\theta _{n - 1}}$ are the nonzero singular values of $D{R^{\frac{1}{2}}}$. Since $D=D_{ST}Q$, we can obtain that $\left[ {\begin{array}{*{20}{c}}
\Sigma &{}\\
{}&0
\end{array}} \right] = {V^T}{R^{\frac{1}{2}}}{D^T}D{R^{\frac{1}{2}}}V = {V^T}{R^{\frac{1}{2}}}{Q^T}D_{ST}^T{D_{ST}}Q{R^{\frac{1}{2}}}V$. Noting $D_{ST} \in \mathbb{R}^{n\times(n-1)}$, it follows from $\text{rank}(D_{ST}^T D_{ST})=\text{rank}(D_{ST}) = n-1$   that $D_{ST}^T D_{ST}\in\mathbb{R}^{(n-1)\times (n-1)}$ is positive definite. 
By inspecting the equation
\[
\left[ {\begin{array}{*{20}{c}}
\Sigma &{}\\
{}&0
\end{array}} \right]  = {V^T}{R^{\frac{1}{2}}}{Q^T}D_{ST}^T{D_{ST}}Q{R^{\frac{1}{2}}}V,
\]
it can be implied that $Q{R^{\frac{1}{2}}}V = \left[ {\begin{array}{*{20}{c}}
U&0
\end{array}} \right]$ for some full-rank $U\in\mathbb{R}^{(n-1)\times (n-1)}$ and ${U^T}{{D_{ST}^T}}D_{ST}U = \Sigma$. Now, we are ready to rewrite $M$ into
\begin{align}\label{eq: M}
M &:=Q\left( {{\gamma}{I_p}R + {R}{D^T}DR} \right){Q^T}\nonumber\\
& = Q{R^{\frac{1}{2}}}\left( {\gamma {I_p} + {R^{\frac{1}{2}}}{D^T}D{R^{\frac{1}{2}}}} \right){R^{\frac{1}{2}}}{Q^T}\nonumber\\
& = Q{R^{\frac{1}{2}}}V\left( {\gamma {I_p} + \left[ {\begin{array}{*{20}{c}}
\Sigma &{}\\
{}&0
\end{array}} \right]} \right){V^T}{R^{\frac{1}{2}}}{Q^T}\nonumber\\
& = \left[ {\begin{array}{*{20}{c}}
U&0
\end{array}} \right]\left( {\gamma {I_p} + \left[ {\begin{array}{*{20}{c}}
\Sigma &{}\\
{}&0
\end{array}} \right]} \right)\left[ {\begin{array}{*{20}{c}}
{{U^T}}\\
0
\end{array}} \right]\nonumber\\
& = U\left( {\gamma {I_{n - 1}} + \Sigma } \right){U^T}.
\end{align}
It follows from \eqref{eq: M} that $M\succ 0$  if $\gamma {I_{n - 1}} + \Sigma  \succ 0$. 
On the other hand, since $\text{rank}(D)=n-1$, the nonzero singular values of $D$ can be ordered in a nonincreasing manner as $\sigma_1\ge \sigma_2 \ge \cdots \ge \sigma_{n-1} > 0$. Noting that $\lambda_2$ is the smallest nonzero eigenvalue of $L=DD^T$, we have $\sigma_{n-1} = \sqrt {{\lambda _2}}$. According to the singular value inequalities in \cite{loyka2015novel}, one has ${s_i}\left( {AB} \right) \ge {s_i}\left( A \right){s_{\text{min}}}\left( B \right)$, where ${s_{\text{min}}}\left( B \right)$ is the smallest singular value of $B$, ${s_i}\left( A \right)$ and ${s_i}\left( AB \right)$ are the $i$th largest singular values of $A$ and $AB$ respectively. Therefore, we can obtain that  ${\theta _{n - 1}} ={s_{n - 1}}\left( {{D}{R^{\frac{1}{2}}}} \right) \ge s_{n-1}\left( {{D}} \right){s_\text{min}}\left( {{R^{\frac{1}{2}}}} \right) = {\sigma _{n - 1}}{s_\text{min}}\left( {{R^{\frac{1}{2}}}} \right) \ge \sqrt {r{\lambda _2}}>0 $, where ${s_\text{min}}\left( {R^{\frac{1}{2}}} \right)$ is the smallest singular value of ${R^{\frac{1}{2}}}$, ${s_{n-1}}\left( D \right)$ and ${s_{n-1}}\left( D{R^{\frac{1}{2}}} \right)$ are the $(n-1)$th largest singular values (i.e., the smallest nonzero singular values) of $D$ and $D{{R^{\frac{1}{2}}}}$ respectively, and the last inequality follows from $r_i \ge r >0$. Accordingly, we can conclude that if $\gamma + r\lambda_2 >0$, then $\gamma + \theta_{n-1}^2 >0$ and thus $\gamma {I_{n - 1}} + \Sigma  \succ 0$, i.e., $M = Q\left( {{\gamma}{I_p}R + {R}{D^T}DR} \right){Q^T}\succ 0$. 
\hfill $\qed$
\end{pf}

For a  homogeneous network (i.e., all nodes share the same dynamics), suppose  that the node dynamics are $\gamma_c$-incremental OFP and the linear diffusive coupling gain is given  by a constant $\alpha$.  It has been shown by \cite{scardovi2010synchronization}  that \eqref{eq:aim} holds if $\gamma_c+ \alpha \lambda_2>0$. In the next theorem, we generalise this result to the case of heterogeneous networks. 
\begin{thm}\label{thm: consensus condition}
Consider the interconnected network \eqref{eq: system model}-\eqref{eq: coupling} and suppose Assumption \ref{assump: 1} holds. Let $\gamma_{m}=\mathop {\min }\limits_{(i,j) \in \mathcal{E}} {\gamma_{ij}}$ and $\alpha = \mathop {\min }\limits_{(i,j) \in \mathcal{E}} {\alpha_{ij}}$.  Then, \eqref{eq:aim} holds if $\gamma_m + \alpha\lambda_2>0$. 
\end{thm}

\begin{pf}
According to Lemma \ref{lem: Q}, there exists a matrix $Q \in \mathbb{R}^{(n-1) \times p}$ such that $D=D_{ST}Q$. Consequently, for all $U \in \Ltwoe$,
\begin{align}\label{eq: DY,DU}
{\left\langle {\Psi {{D^T}Y},{D^T}U} \right\rangle _T}
&= \frac{1}{2}\sum\limits_{(i,j) \in \mathcal{E}} {{\alpha_{ij}}{{\left\langle {{{y_i} - {y_j}} ,{u_i} - {u_j}} \right\rangle }_T}}  \nonumber\\
& \ge \frac{1}{2}\sum\limits_{(i,j) \in \mathcal{E}} {{\alpha _{ij}}\left( {{\gamma _{ij}}\left\| {{y_i} - {y_j}} \right\|_T^2 + {\beta _{ij}}} \right)}  \nonumber\\
& \ge {\left\langle {{D^T}Y,{\gamma _m}I_p\Psi {D^T}Y} \right\rangle _T}+\bar\beta \nonumber\\
&= {\left\langle {D_{ST}^TY,Q{\gamma _m}I_p\Psi {Q^T}D_{ST}^TY} \right\rangle _T}+\bar\beta,
\end{align}
where $\bar\beta= \frac{1}{2}\sum\limits_{(i,j) \in \mathcal{E}} {{\alpha_{ij}}\beta_{ij}} $, the first inequality follows from Assumption \ref{assump: 1}, and $\Psi$ is defined after \eqref{eq: iuput vector}. On the other hand, we have that
\begin{align}\label{eq: DY,DV}
{\left\langle {\Psi {{D^T}Y},{D^T}V} \right\rangle _T}
& = \int_0^T {{Y^T}D\Psi {D^T}D\Psi {D^T}Ydt}  \nonumber \\
& = \int_0^T {{Y^T}{D_{ST}}Q\Psi {D^T}D\Psi {Q^T}D_{ST}^TYdt} \nonumber \\
& = {\hspace{-0.4mm}}{\left\langle {\hspace{-0.4mm}}{D_{ST}^TY,Q\Psi {D^T}D\Psi {Q^T}D_{ST}^TY}{\hspace{-0.4mm}} \right\rangle _T}.
\end{align}
Define ${\tilde M}:=Q\left( {{\gamma _m}{I_p}\Psi  + \Psi {D^T}D\Psi } \right){Q^T}$. By hypothesis, $\gamma_m +\alpha \lambda_2>0$, and thus according to Lemma \ref{lem: positive definite}, ${\tilde M}\succ 0$, leading to
\begin{align}\label{eq: OFP+graph}
{\left\langle {\Psi {{D^T}Y},{D^T}W} \right\rangle _T}
&= {\left\langle {\Psi  {{D^T}Y} ,{D^T}U} \right\rangle _T}+ {\left\langle {\Psi {{D^T}Y},{D^T}V} \right\rangle _T}\nonumber\\
&  \ge \int_0^T {{Y^T} D_{ST} \tilde M{{D_{ST}^T}}Y}dt +\bar\beta \nonumber\\ 
& \ge \mu \left\| {{{D_{ST}^T}}Y} \right\|_T^2+\bar\beta,
\end{align}
where $\mu$ is the smallest eigenvalue of $\tilde M$. Note that $\left\| {\Psi {{D^T}Y}} \right\|_T^2 {\hspace{-0.4mm}}\le{\hspace{-0.4mm}} \bar \alpha^2 \left\| {{D^T}Y} \right\|_T^2$ and $\left\| {{D^T}Y} \right\|_T^2 \le \kappa \left\| {D_{ST}^TY} \right\|_T^2$, where $\bar \alpha = \mathop {\max }\limits_{(i,j) \in \mathcal{E}} {\alpha_{ij}}$ and $\kappa$ is the largest eigenvalue of $QQ^T$. Thus, we obtain from \eqref{eq: OFP+graph} that
\begin{align*}
\frac{\mu }{\kappa }\left\| {{D^T}Y} \right\|_T^2 &\le \mu \left\| {D_{ST}^TY} \right\|_T^2 \le {\left\langle {\Psi {D^T}Y,{D^T}W} \right\rangle _T} - \bar \beta \\
& \le {\left\langle {\Psi {D^T}Y,{D^T}W} \right\rangle _T} - \bar \beta \\
&\quad+ \frac{1}{2}\left\| {\sqrt {\frac{\mu }{{\kappa {{\bar \alpha }^2}}}} \Psi {D^T}Y - \sqrt {\frac{{\kappa {{\bar \alpha }^2}}}{\mu }} {D^T}W} \right\|_T^2\\
& = \frac{\mu }{{2\kappa {{\bar \alpha }^2}}}\left\| {\Psi {D^T}Y} \right\|_T^2 + \frac{{\kappa {{\bar \alpha }^2}}}{{2\mu }}\left\| {{D^T}W} \right\|_T^2 - \bar \beta \\
& \le \frac{\mu }{{2\kappa }}\left\| {{D^T}Y} \right\|_T^2 + \frac{{\kappa {{\bar \alpha }^2}}}{{2\mu }}\left\| {{D^T}W} \right\|_T^2 - \bar \beta. 
\end{align*}
This implies
\begin{align}\label{eq: DY,DW}
\left\| {{D^T}Y} \right\|_T^2 \le \frac{{{\kappa ^2}{{\bar \alpha }^2}}}{{{\mu ^2}}}\left\| {{D^T}W} \right\|_T^2 - \frac{{2\kappa \bar \beta }}{\mu }.    
\end{align}
It follows from \eqref{eq: DY,DW} and ${a^2} \pm {b^2} \le  {\left( {\left| a \right| + \left| b \right|} \right)^2}$ that 
$${\left\| {D^T Y} \right\|_T}\le \rho{\left\| {D^T W} \right\|_T}+\varepsilon,\,\forall {W } \in \Ltwoe,\,\forall T \ge 0,$$
where $\rho  = \frac{{\kappa \bar \alpha }}{\mu }>0 $ and $\varepsilon  = \sqrt {\frac{{2\kappa \left| {\bar \beta } \right|}}{\mu }}\ge 0 $. \hfill $\qed$
\end{pf}

\section{Conclusion}
This paper investigated the consensus problem for networks of heterogeneous agents with external disturbance and/or reference input over diffusive coupling. We introduced the indices that characterise the gaps between the adjacent subsystems. It has been shown that the output of the subsystems in the heterogeneous network reach a certain level of consensus if  the sum of the level of heterogeneity of the network and the connectivity of the communication graph is positive. 
\bibliography{ifacconf}
\end{document}